\documentclass[floatfix,pre,showpacs,twocolumn]{revtex4}
\usepackage{graphicx}
\usepackage{amsmath}
\usepackage{bm}
\usepackage{latexsym}
\usepackage{verbatim}
\begin{document}

\title{Nematic topological defects in the presence of axisymmetric fluid flow}

\author{Robert~A.~Pelcovits and Pengyu~Liu}
\affiliation{Department of Physics, Brown University, Providence,
Rhode Island 02912}
\date{\today}
\begin{abstract}
Recent numerical simulations of lid--driven cavity flow of a nematic liquid crystal have found dynamical behavior where topological defects rotate about the center of the fluid vortex induced by the lid motion. By considering a simpler geometry of an infinite system with axisymmetric fluid flow we show that the Ericksen--Leslie nematodynamic equation for the director can be solved exactly. The solution demonstrates that any configuration of defects will be advected by the fluid flow, with the defects rotating about the center of the fluid vortex with the angular velocity of the fluid.
\end{abstract}
\maketitle
Lid--driven cavity flow is a standard geometry used in both experimental and computational studies of fluids. In this geometry a fluid is contained within a box whose sides are stationary except for the the top wall (the ``lid") which is driven at some nonzero velocity. In numerical studies the problem is often simplified to a two--dimensional square domain with one moving side. The driven lid creates a vortex fluid flow near the center of the cavity as well as counterrotating vortices near the corners. If the fluid is a complex one, such as a nematic liquid crystal, the driven lid can lead to very interesting and rich topological defect phenomena as has been observed recently in a number of numerical simulation studies \cite{Yang:2010,Hernandez:2011,Liu:2012}. Among the phenomena observed are rotating half--integer charged topological defects, either a single defect \cite{Hernandez:2011} or a pair of oppositely charged defects \cite{Yang:2010} ,depending on the director boundary conditions employed. The defects are advected by the fluid vortex near the center of the cavity and undergo a periodic rotation with the fluid flow about the center of the vortex. In the case of a single defect \cite{Hernandez:2011} the defect core eventually settles at the center of the vortex and remains stationary, while in the case of a defect pair \cite{Yang:2010} the defects continue to rotate indefinitely with other transient defect phenomena occuring elsewhere in the cavity.

In this paper we show that by considering a simpler but related geometry the Ericksen--Leslie nematodynamic equation for the director can be solved exactly for any texture composed of an array of defects. The solution is consistent with the behavior observed in the simulations, namely, advection of the defects about the center of the fluid vortex with an angular speed equal to that of the fluid.

Specifically, we consider a nematic liquid crystal with a simple vortex fluid flow (a ``stationary axisymmetric flow") given in plane polar coordinates by:
\begin{equation}
\label{vortex}
\mathbf{v} = \omega r\mathbf{e}_\phi
\end{equation}
where $r$ measures the radial distance from the center of the vortex and $\mathbf{e}_\phi$ is the tangential polar unit vector.
The angular speed of the fluid $\omega$ is the magnitude of the angular velocity:
\begin{equation}
\mathbf{\omega}=\frac{1}{2}\nabla \times \mathbf{v}.
\label{eq:D_6}
\end{equation}
We assume that the system is infinite in size, thus simplifying the cavity flow geometry while retaining the essence of the lid--driven flow, namely, the central vortex. 
For this vortex flow the symmetric velocity tensor $d_{ij}=\frac{1}{2}\left(\frac{\partial v_{i}}{\partial x_{j}}+\frac{\partial v_{j}}{\partial x_{i}}\right)$
is identically zero and the Ericksen--Leslie hydrodynamic equation for the nematic director \cite{Stephen:74} simplifies to:
\begin{equation}
\gamma\mathbf{n}=K\nabla^{2}\mathbf{n}+(\alpha_{3}-\alpha_{2})\mathbf{N}
\label{eq:D_1}
\end{equation}
assuming a single Frank elastic constant $K$. 
Here $\alpha_{2},\alpha_{3}$  are Leslie viscosity coefficients and $\mathbf{N}$ is the rotation rate of the director relative to the moving fluid:
\begin{equation}
\mathbf{N}=\frac{\partial}{\partial t}\mathbf{n} + \mathbf{v}\cdot\nabla\mathbf{n}-\mathbf{\omega}\times\mathbf{n}
\label{eq:D_4}
\end{equation}
The Lagrange multiplier $\gamma$ maintains the constraint:  $| \mathbf{n}| \equiv 1$.

We assume that the director field lies in the plane of the fluid flow (the $x-y$ plane) and express it in rectangular coordinates by:
\begin{equation}
\label{n}
 \mathbf{ n } =(\cos\theta (x, y), \sin \theta(x, y))
 \end{equation}
 where $\theta$ is the angle $\mathbf{n}$ makes with the $x$ axis.  Substituting Eqs.~(\ref{vortex}) and (\ref{n}) into Eq.~(\ref{eq:D_1}) the two nonzero components of the latter equation are found to be:
\begin{eqnarray}
\gamma\cos\theta & = &A \sin\theta  \left(-\frac{\partial\theta}{\partial t} +\omega y \frac{\partial\theta}{\partial x} - \omega x \frac{\partial\theta}{\partial y}\right)\nonumber\\
 && -K \left[-\cos\theta\left( \left(\frac{\partial \theta}{\partial x}\right)^2 + \left(\frac{\partial \theta}{\partial y}\right)^2 \right)\right.\nonumber\\
  &&- \left.\sin\theta \left( \frac{\partial ^2\theta}{\partial x^2}+ \frac{\partial ^{2}\theta}{\partial y^{2}} \right)\right]+A \omega\sin\theta\\
\label{eq:D_8a}
\gamma\sin\theta & = &-A \cos\theta  \left(-\frac{\partial\theta}{\partial t} +\omega y \frac{\partial\theta}{\partial x} - \omega x \frac{\partial\theta}{\partial y}\right)\nonumber\\
 && -K \left[-\sin\theta\left( \left(\frac{\partial \theta}{\partial x}\right)^2 + \left(\frac{\partial \theta}{\partial y}\right)^2 \right)\right.\nonumber\\
  &&+ \left.\cos\theta \left( \frac{\partial ^2\theta}{\partial x^2}+ \frac{\partial ^{2}\theta}{\partial y^{2}} \right)\right]-A \omega\cos\theta\\
\label{eq:D_8b}
\end{eqnarray}
where $A=\alpha_3-\alpha_2$.
After eliminating the unknown Lagrange multiplier $\gamma$ from the above two equations we obtain:
\begin{equation}
\frac{\partial}{\partial t}\theta+\omega\left(x\frac{\partial}{\partial y}-y\frac{\partial}{\partial x}\right)\theta -\omega
=\frac{K}{A}\nabla^{2}\theta. \label{eq:D_9}
\end{equation}
The left hand side of Eq.~(\ref{eq:D_9}) is the time rate of change of the director angle $\theta$ relative to the rotating fluid \cite{rotational_derivative}.

If initially $\nabla ^2 \theta=0$, a condition satisfied at all points in space by any configuration of topological defects except at the isolated defect cores \cite{deGennes:93}, then Eq.~(\ref{eq:D_9}) requires that the time rate of change of $\theta$ relative to the rotating fluid is zero, i.e., $\theta$ as a function of the polar coordinates in the nonrotating frame must satisfy:
\begin{equation}
\theta(r,\phi,t)=\theta(r,\phi - \omega\Delta t,t-\Delta t)+\omega\Delta t
\label{eq:D_13}
\end{equation}
for all $t$ and $\Delta t$ Thus, if there are topological defects in the system initially the director pattern will rotate with the fluid, i.e., the defects will rotate with same angular velocity $\mathbf{\omega}$ as the fluid. The above equation is the exact solution to the Ericksen--Leslie director equation for an axisymmetric flow and an initial director configuration satisfying $\nabla ^2 \theta=0$. While the latter equation is not satisfied at the defect core itself, the core must rotate at the angular velocity as the director field surrounding it to preserve the continuity of the structure. Also, note that this solution is stable against director fluctuations out of the plane as can be seen from looking at the $z$ component of Eqn.~(\ref{eq:D_1}).

\begin{figure*}
\begin{footnotesize}
\begin{tabular}{cc}

\includegraphics[scale=0.35]{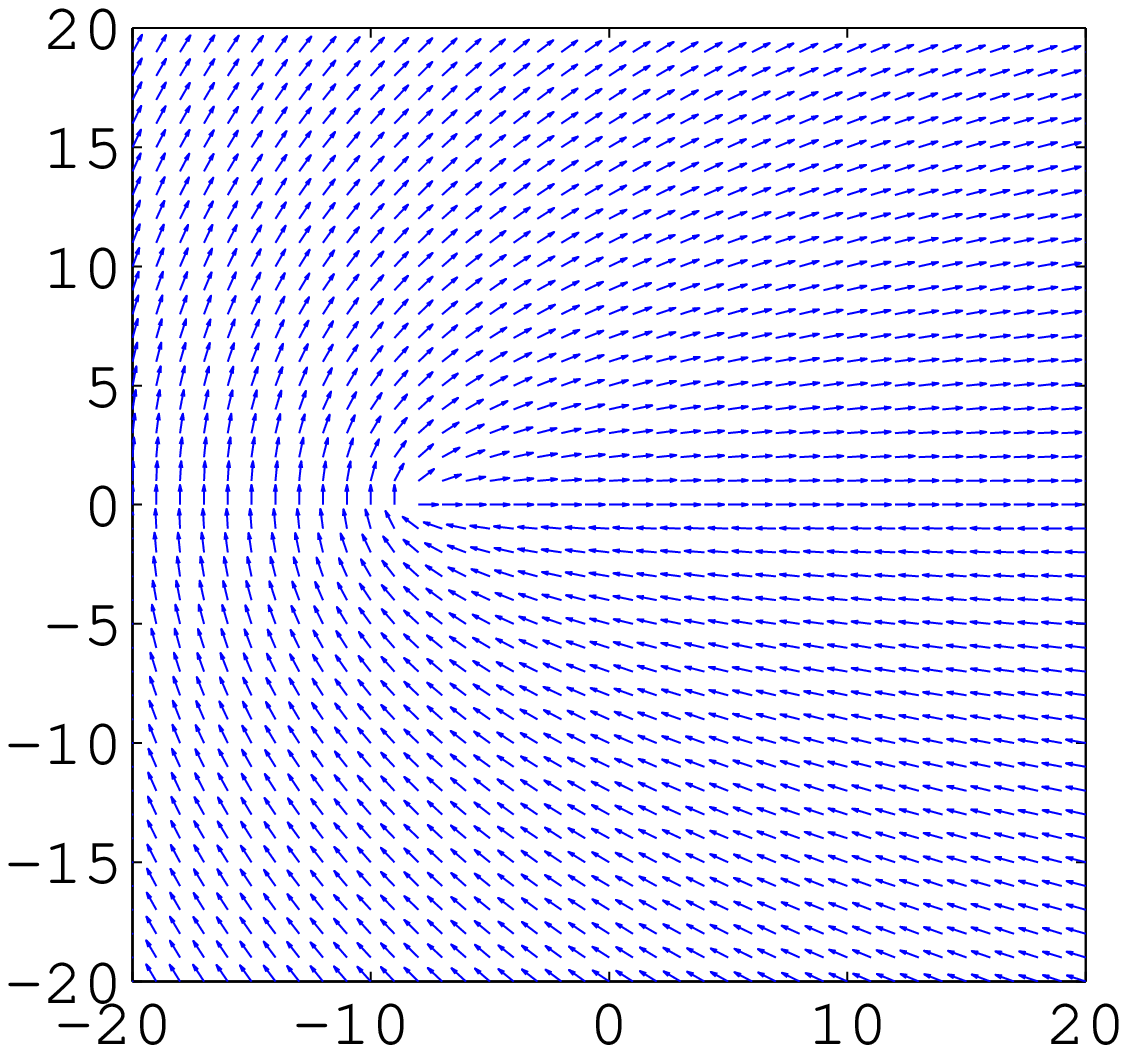} & \includegraphics[scale=0.35]{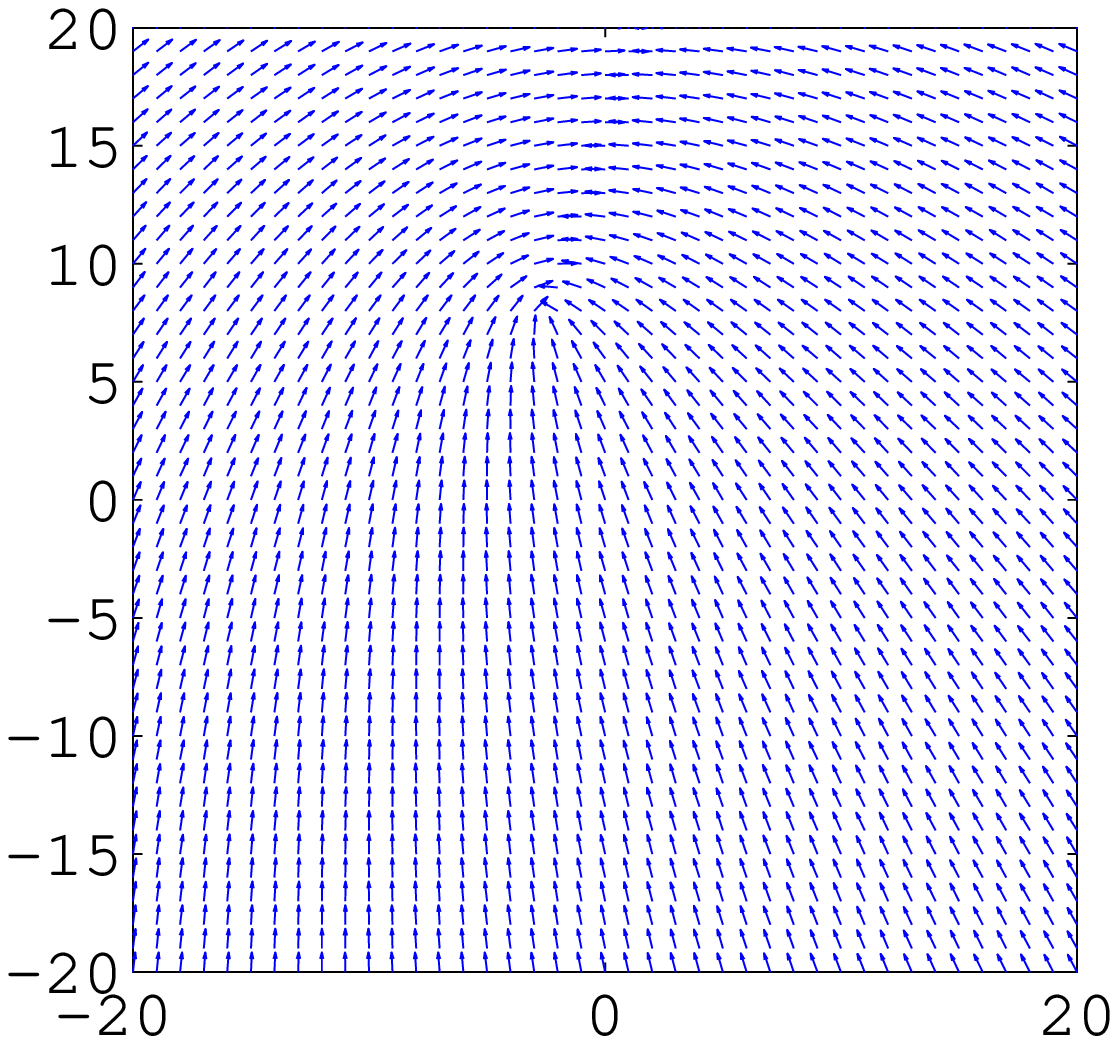}\\
(a) $t=0.0 $ & (b) $t=0.4 $\\
\includegraphics[scale=0.35]{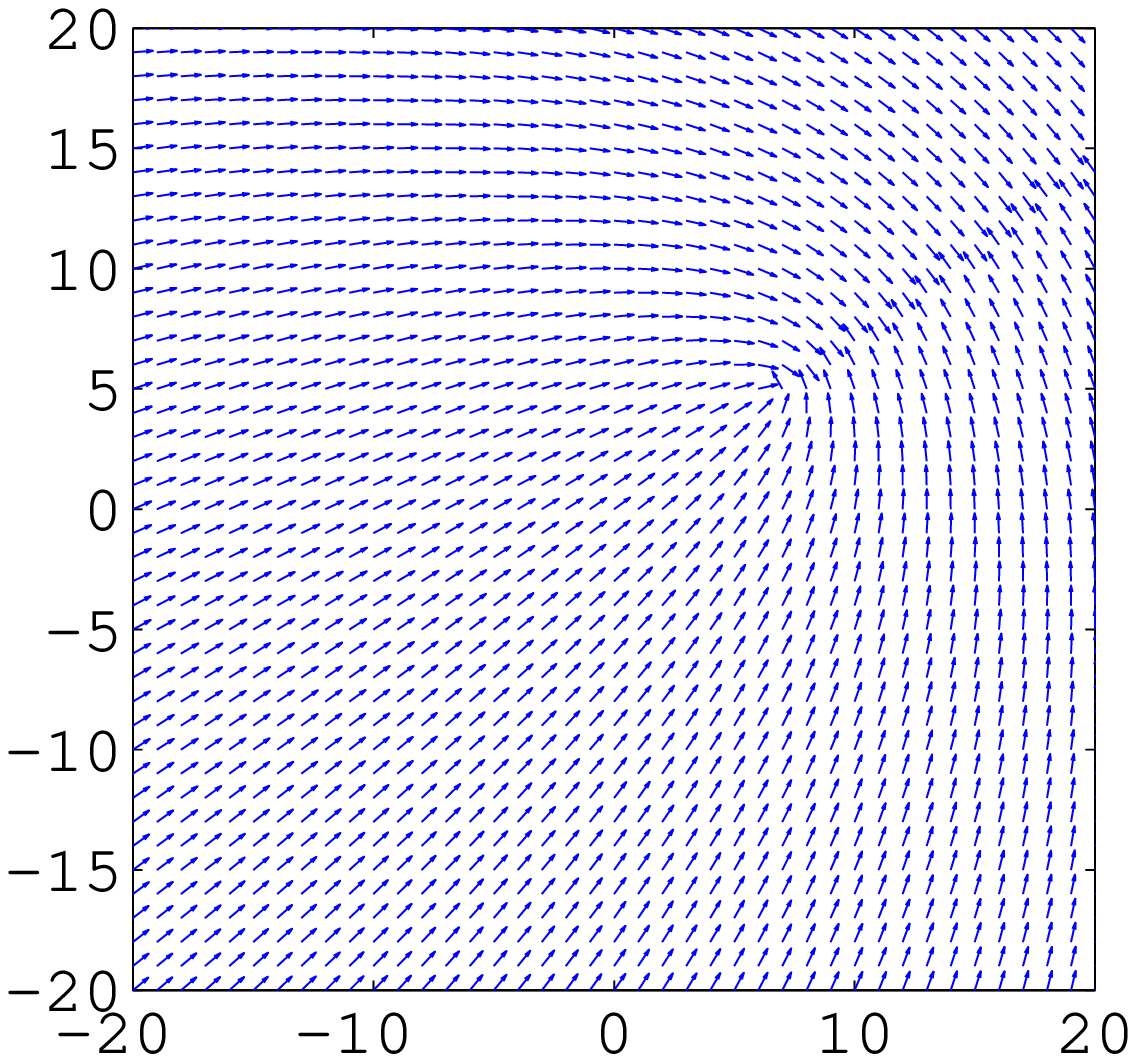} & \includegraphics[scale=0.35]{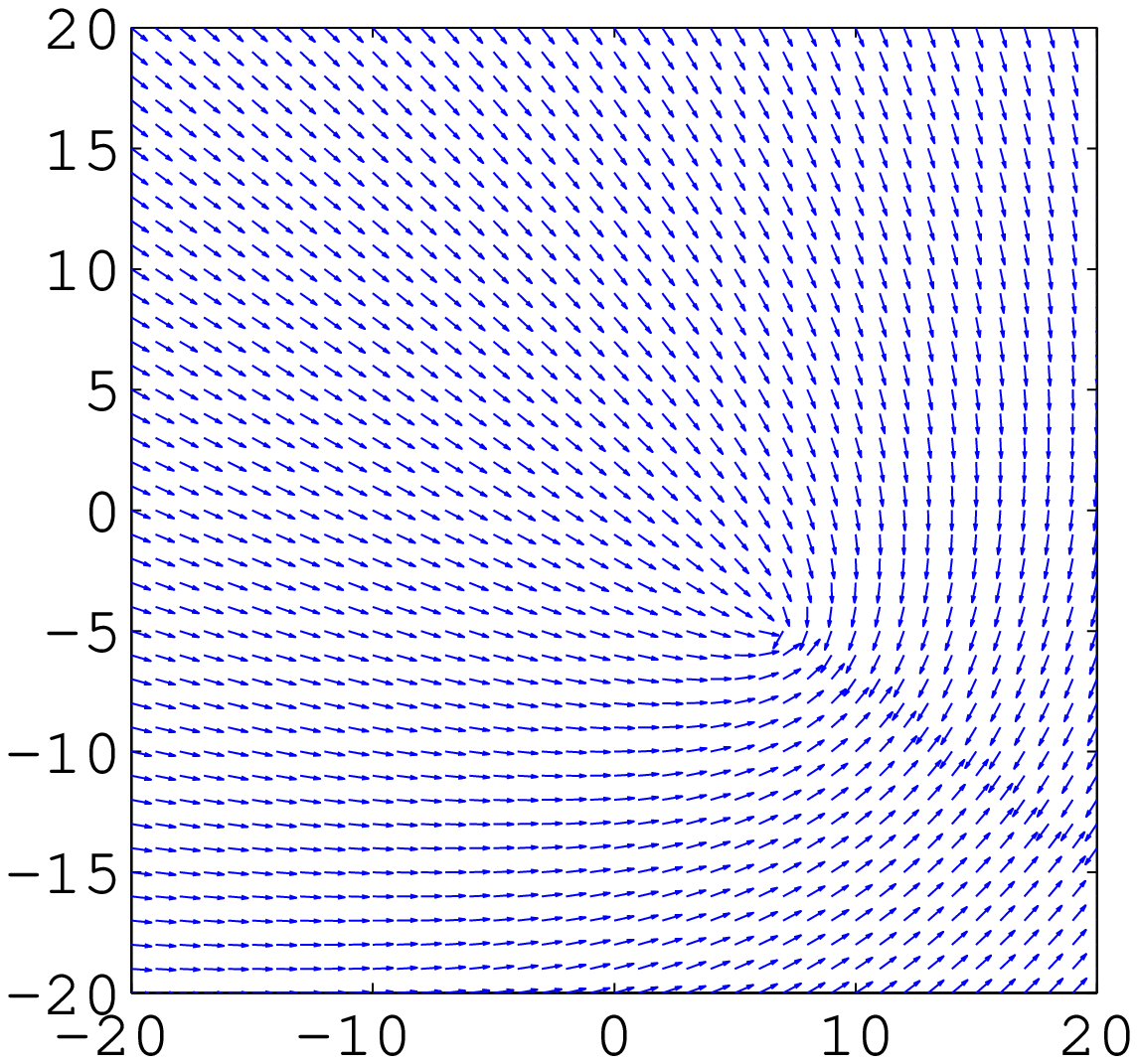}\\
(c) $t=0.8 $ & (d) $t=1.2 $\\
\includegraphics[scale=0.35]{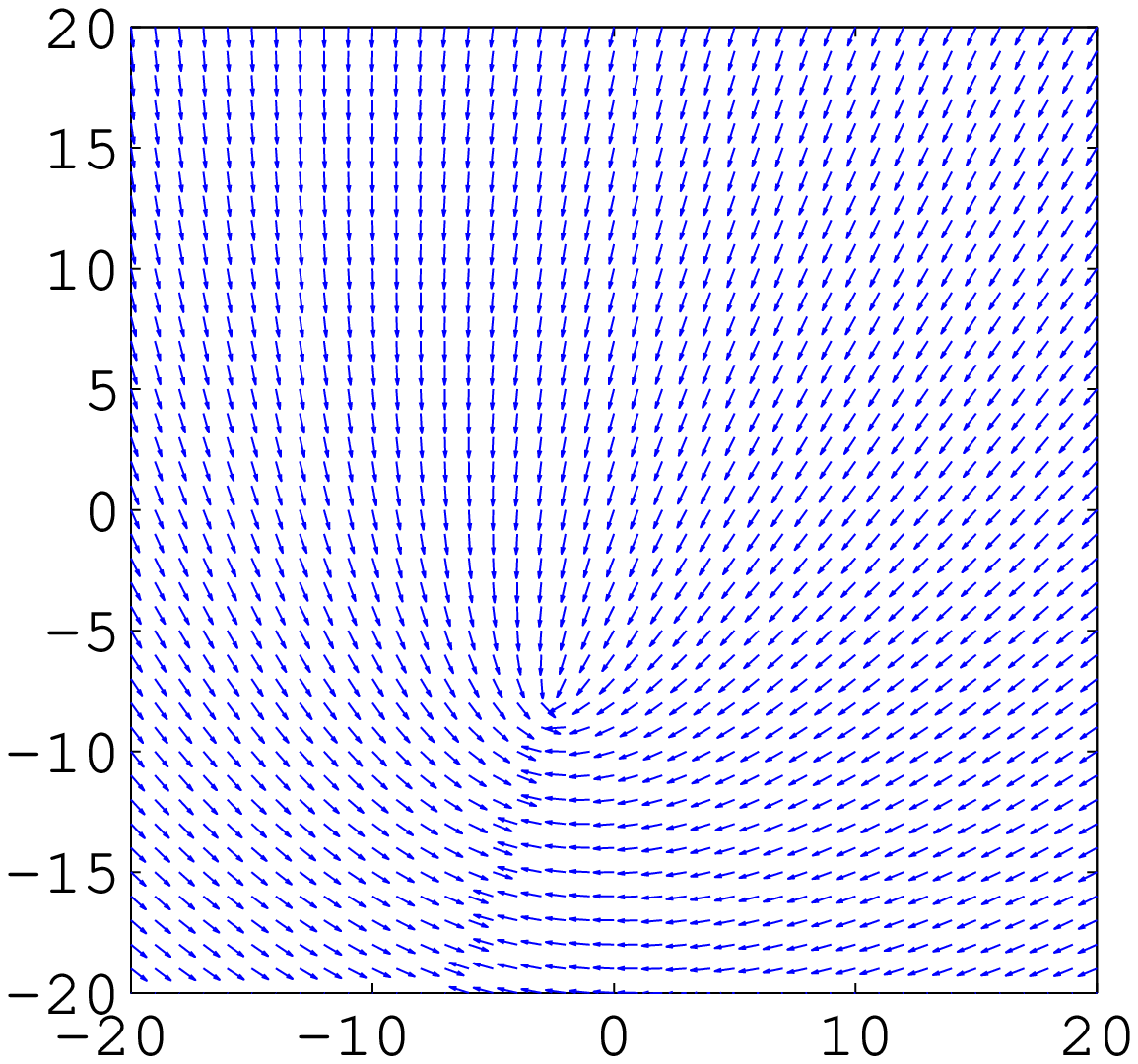} & \includegraphics[scale=0.35]{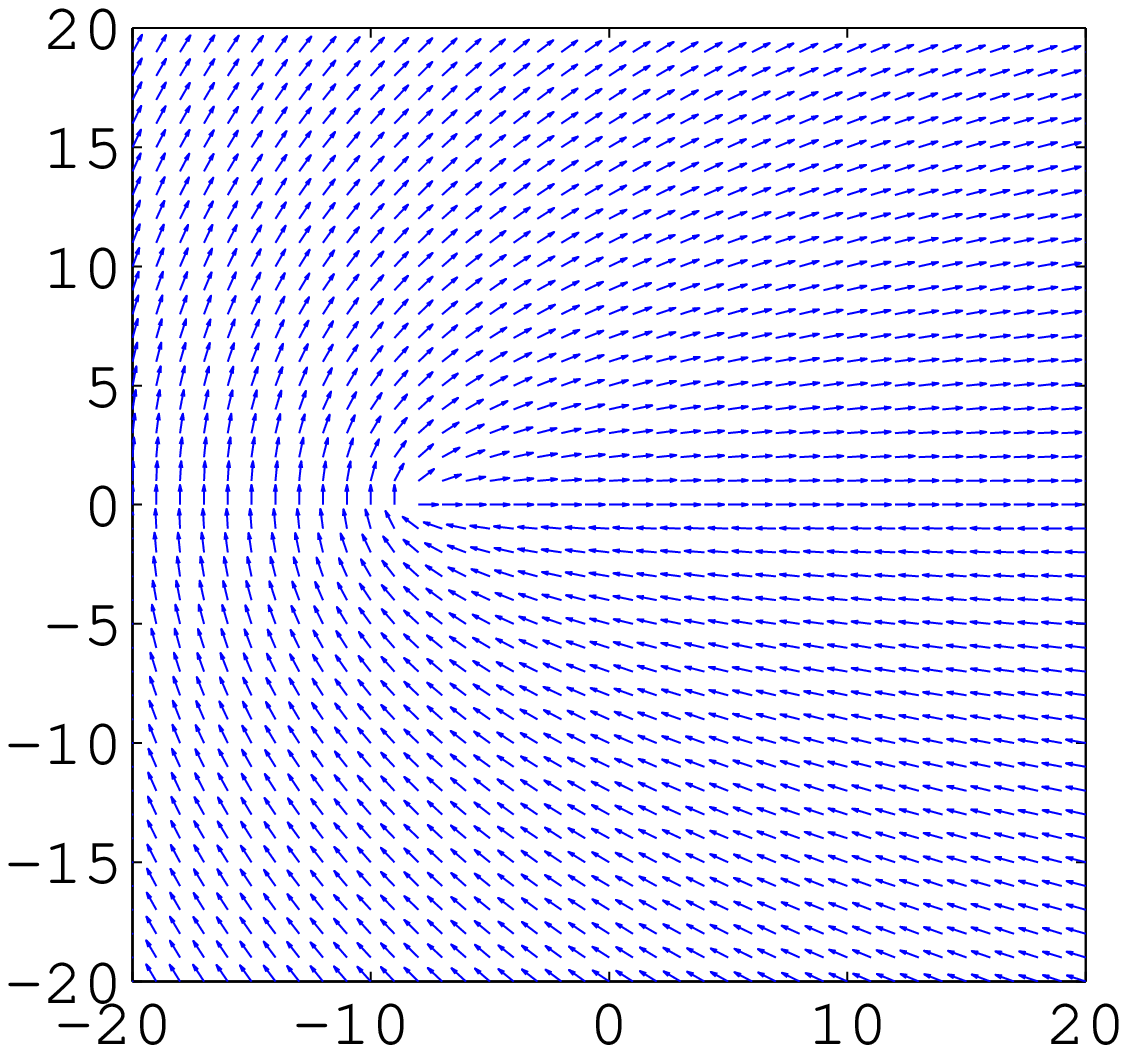}\\
(e) $t=1.6 $ & (f) $t=2.0 $\\
\end{tabular}
\end{footnotesize}
\caption{Illustration of the motion of a single half--integer topological defect with the director angle $\theta$ given by Eq.~(\ref{eq:D_13}). We have chosen $ \omega=-\pi$, i.e, a clockwise fluid flow and placed the defect initially at  $x=-4.3$. Compare with Fig.~5 of \cite{Hernandez:2011}.}
\label{fig:SingleHalfMotion}
\end{figure*}

\begin{figure*}
\begin{tabular}{cc}
\includegraphics[scale=0.35]{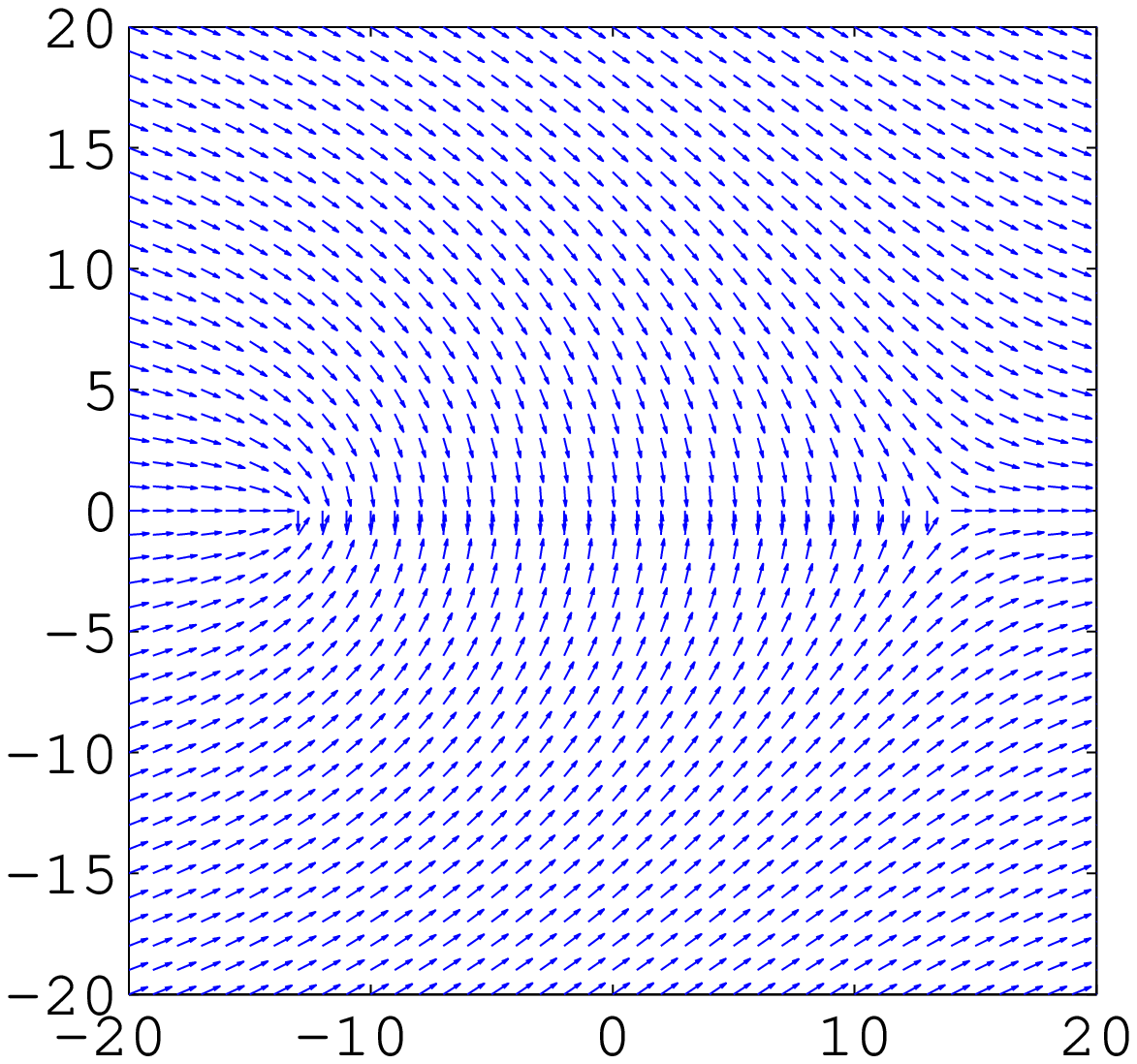} & \includegraphics[scale=0.35]{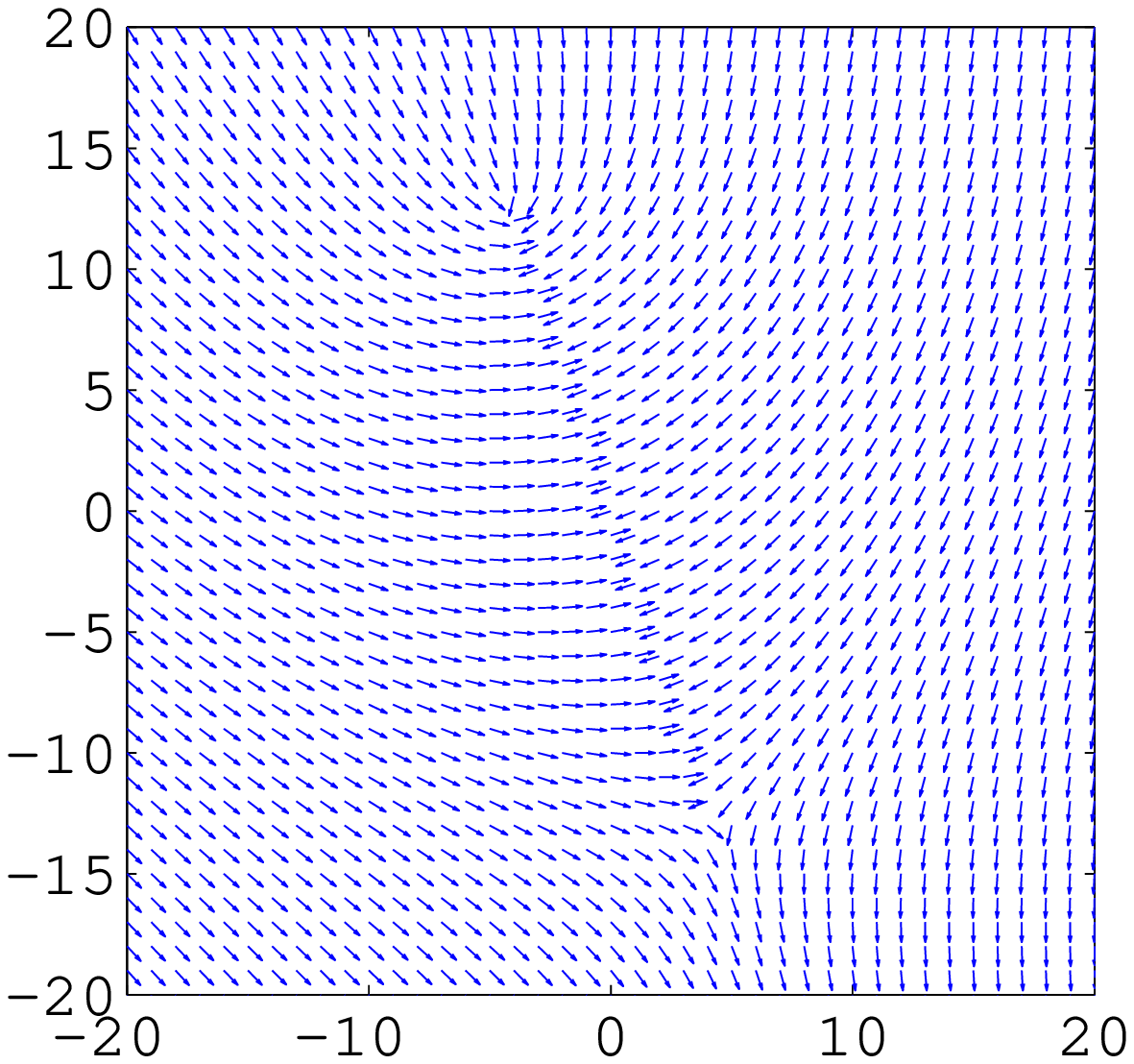}\\
(a) $t=0.0 $ & (b) $t=0.4 $\\
\includegraphics[scale=0.35]{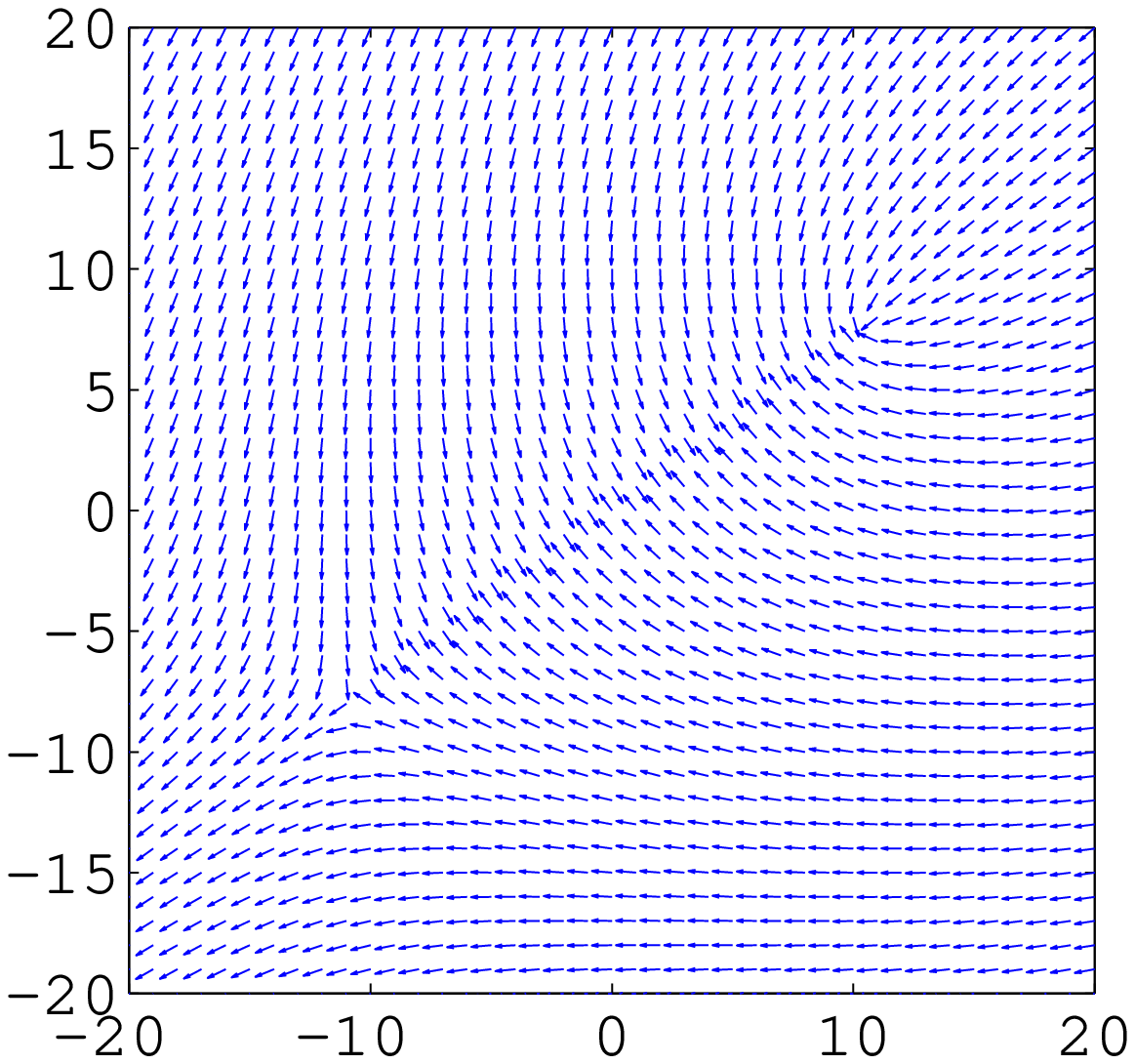} & \includegraphics[scale=0.35]{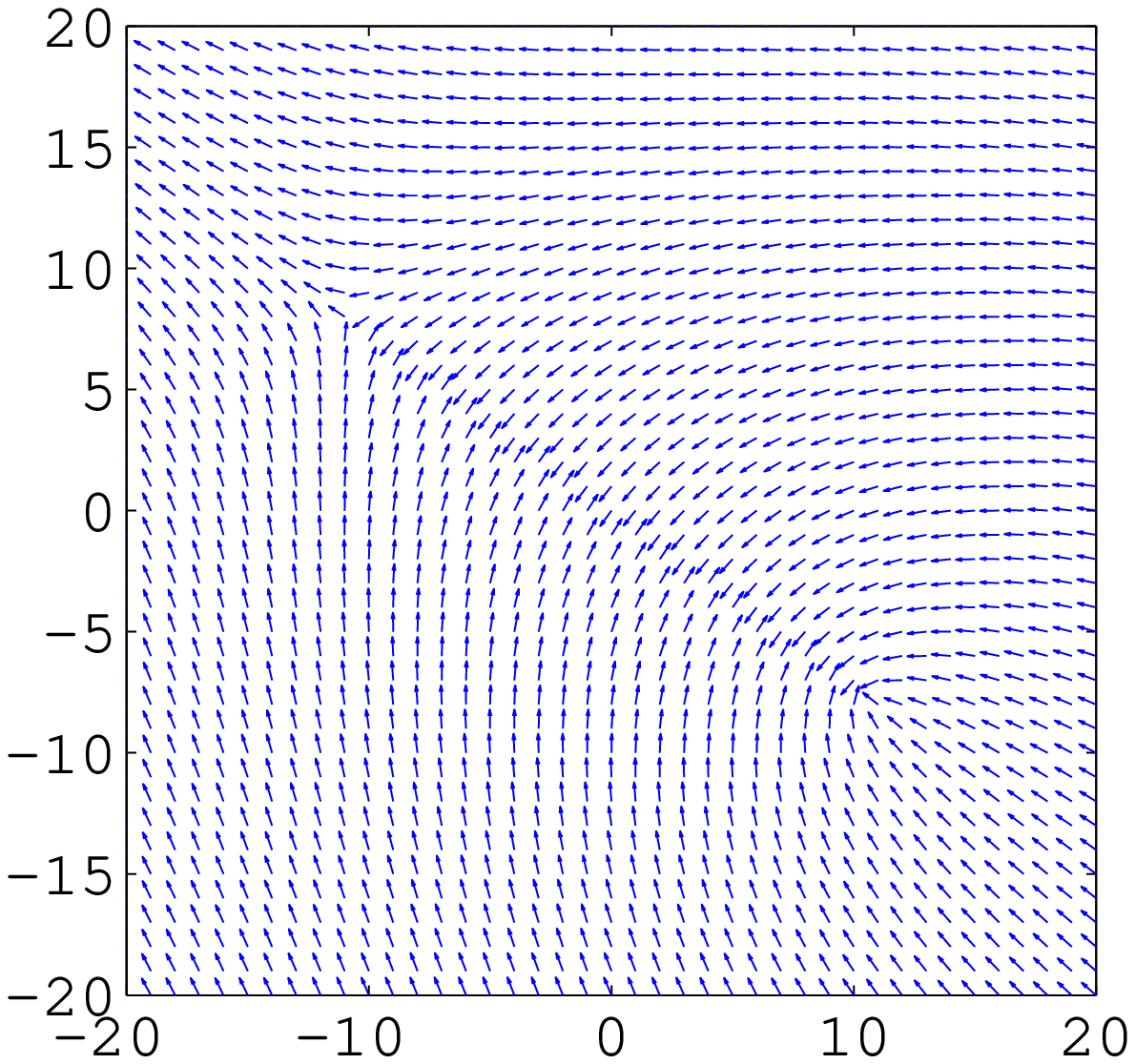}\\
(c) $t=0.8 $ & (d) $t=1.2 $\\
\includegraphics[scale=0.35]{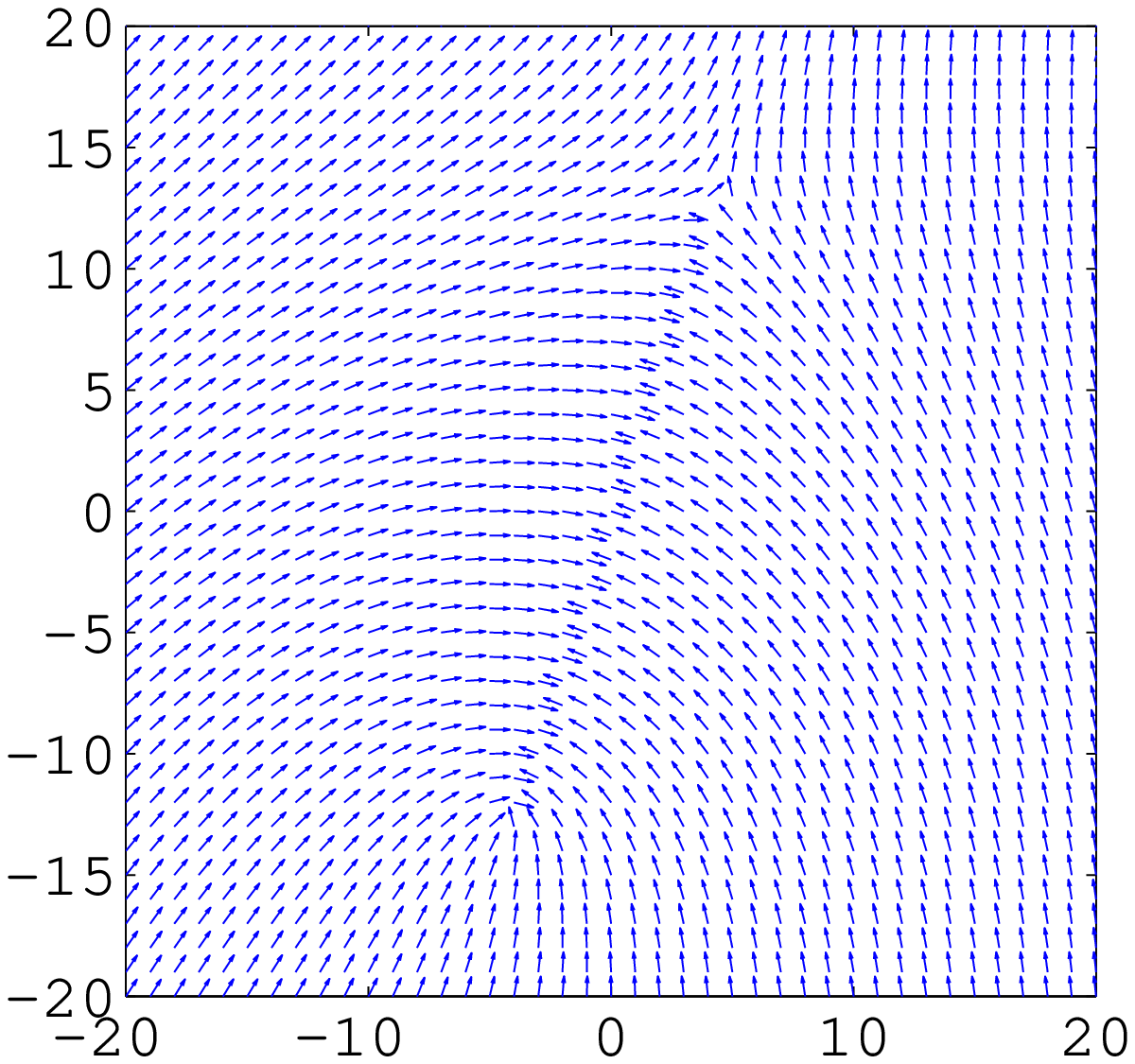} & \includegraphics[scale=0.35]{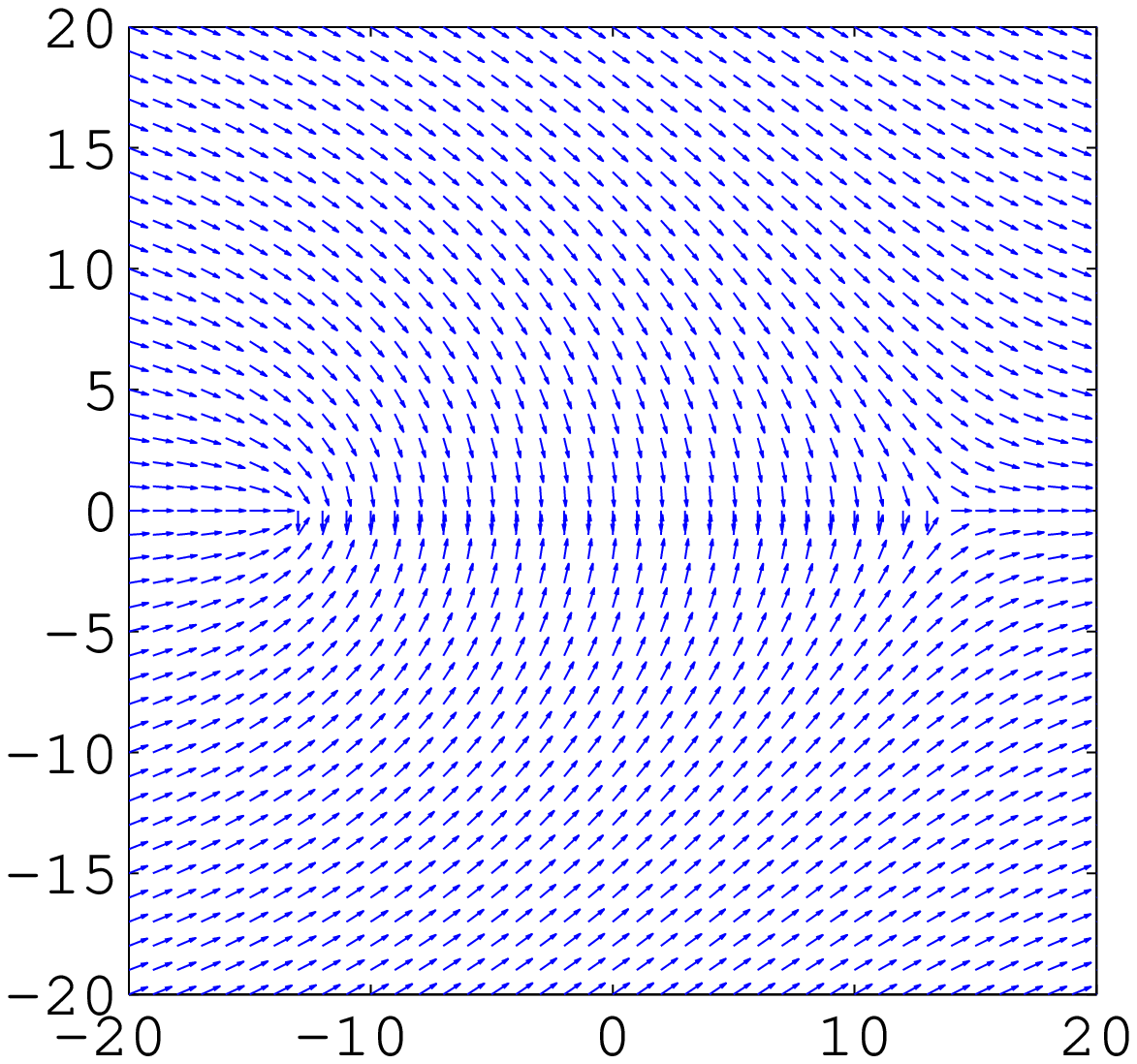}\\
(e) $t=1.6 $ & (f) $t=2.0 $\\
\end{tabular}

\caption{Illustration of the motion of a pair of oppositely charged half--integer topological defects with the director angle $\theta$ given by Eq.~(\ref{eq:D_13}). We have chosen $ \omega=-\pi$, i.e, a clockwise fluid flow and placed the defects initially at a distance 13.5 on either side of the origin. Compare with Fig.~8 of \cite{Yang:2010}.}
\label{fig:HalfPairMotion}
\end{figure*}

In Figs.~\ref{fig:SingleHalfMotion} and \ref{fig:HalfPairMotion} we illustrate Eq.~(\ref{eq:D_13}) for two cases: a single half--integer defect and a pair of oppositely charged half--integer defects, configurations that were observed in the numerical simulations of lid--driven cavity flow in Refs.~\cite{Hernandez:2011} and \cite{Yang:2010}, respectively. In both figures we have chosen $\omega=-\pi$, where the negative sign corresponds to the clockwise fluid flow imposed in the simulations by a lid driven in the positive $x$ direction. The figures illustrate the periodic rotation of the defects with period $2\pi/\omega =2 $ as observed in the numerical simulations. This is the steady--state motion for the axisymmetric infinite system considered in this paper.  In the simulation of Ref.~\cite{Hernandez:2011} the single defect eventually settled to the center of the vortex; this behavior arises presumably from the presence of the defects at the corners of the cavity leading to an equilibrium position of the rotating defect at the vortex center. In the case of the rotating pair of defects \cite{Yang:2010} the rotation was observed to occur indefinitely as is the case for our simple geometry.

\section*{Acknowledgments}
We are grateful to J.~X. Tang for helpful discussions. Acknowledgment is made to the Donors of the American Chemical Society Petroleum Research Fund for partial support of this research.


\begin{thebibliography}{6}
\expandafter\ifx\csname natexlab\endcsname\relax\def\natexlab#1{#1}\fi
\expandafter\ifx\csname bibnamefont\endcsname\relax
  \def\bibnamefont#1{#1}\fi
\expandafter\ifx\csname bibfnamefont\endcsname\relax
  \def\bibfnamefont#1{#1}\fi
\expandafter\ifx\csname citenamefont\endcsname\relax
  \def\citenamefont#1{#1}\fi
\expandafter\ifx\csname url\endcsname\relax
  \def\url#1{\texttt{#1}}\fi
\expandafter\ifx\csname urlprefix\endcsname\relax\def\urlprefix{URL }\fi
\providecommand{\bibinfo}[2]{#2}
\providecommand{\eprint}[2][]{\url{#2}}

\bibitem[{\citenamefont{Yang et~al.}(2010)\citenamefont{Yang, Forest, Mullins,
  and Wang}}]{Yang:2010}
\bibinfo{author}{\bibfnamefont{X.}~\bibnamefont{Yang}},
  \bibinfo{author}{\bibfnamefont{M.~G.} \bibnamefont{Forest}},
  \bibinfo{author}{\bibfnamefont{W.}~\bibnamefont{Mullins}}, \bibnamefont{and}
  \bibinfo{author}{\bibfnamefont{Q.}~\bibnamefont{Wang}},
  \bibinfo{journal}{Soft Matter} \textbf{\bibinfo{volume}{6}},
  \bibinfo{pages}{1138} (\bibinfo{year}{2010}).

\bibitem[{\citenamefont{Hern\'andez-Ortiz
  et~al.}(2010)\citenamefont{Hern\'andez-Ortiz, Gettelfinger, Moreno-Razo, and
  de~Pablo}}]{Hernandez:2011}
\bibinfo{author}{\bibfnamefont{J.~P.} \bibnamefont{Hern\'andez-Ortiz}},
  \bibinfo{author}{\bibfnamefont{B.~T.} \bibnamefont{Gettelfinger}},
  \bibinfo{author}{\bibfnamefont{J.}~\bibnamefont{Moreno-Razo}},
  \bibnamefont{and} \bibinfo{author}{\bibfnamefont{J.~J.}
  \bibnamefont{de~Pablo}}, \bibinfo{journal}{J.~Chem.~Phys.}
  \textbf{\bibinfo{volume}{134}}, \bibinfo{pages}{134905}
  (\bibinfo{year}{2010}).

\bibitem[{\citenamefont{Liu}(2012)}]{Liu:2012}
\bibinfo{author}{\bibfnamefont{P.}~\bibnamefont{Liu}}, Ph.D. thesis,
  \bibinfo{school}{Brown University} (\bibinfo{year}{2012}).

\bibitem[{\citenamefont{Stephen and Straley}(1974)}]{Stephen:74}
\bibinfo{author}{\bibfnamefont{M.~J.} \bibnamefont{Stephen}} \bibnamefont{and}
  \bibinfo{author}{\bibfnamefont{J.~P.} \bibnamefont{Straley}},
  \bibinfo{journal}{Rev.~Mod.~Phys.} \textbf{\bibinfo{volume}{46}},
  \bibinfo{pages}{617} (\bibinfo{year}{1974}).

\bibitem[{rot()}]{rotational_derivative}
\bibinfo{note}{The left hand side of Eqn.~(\ref{eq:D_9}) is the
  ''co--rotational derivative'' of $\theta$, just as $\mathbf{N}$ (see
  Eqn.~(\ref{eq:D_4}) is the co--rotational derivative of the director
  $\mathbf{n}$,}.

\bibitem[{\citenamefont{\lowercase{d}e Gennes and Prost}(1993)}]{deGennes:93}
\bibinfo{author}{\bibfnamefont{P.~G.} \bibnamefont{\lowercase{d}e Gennes}}
  \bibnamefont{and} \bibinfo{author}{\bibfnamefont{J.}~\bibnamefont{Prost}},
  \emph{\bibinfo{title}{The Physics of Liquid Crystals}}
  (\bibinfo{publisher}{Clarendon Press, Oxford}, \bibinfo{year}{1993}).

\end{thebibliography}

\end{document}